\newif\ifportablearticle
\IfFileExists{jpconf.cls}{\portablearticlefalse}{\portablearticletrue}
\ifportablearticle
  \documentclass[a4paper,11pt]{article}
  \usepackage[margin=25mm]{geometry}
  \newcommand{\address}[1]{\gdef\paperaddress{#1}}
  \newcommand{\ead}[1]{\gdef\paperemail{#1}}
\else
  \documentclass[a4paper]{jpconf}
\fi
\usepackage{amsmath,amssymb}
\IfFileExists{ubbold.fd}{\usepackage{mathbbol}}{\newcommand{\Eins}{\mathbf{1}}}
\def\bq{\pmb{q}}
\def\bp{\pmb{p}}
\def\bQ{\pmb{Q}}
\def\bP{\pmb{P}}
\def\e{\mathrm{e}}
\def\dd{\mathrm{d}}
\def\onehalf{{\textstyle\frac{1}{2}}}
\def\pfrac#1#2{\frac{\partial #1}{\partial #2}}
\def\dfrac#1#2{\frac{\dd #1}{\dd #2}}
\def\RB{\mathbb{R}}
\def\MB{\mathbb{M}}
\def\AB{\mathbb{A}}
\def\Hv{e}
\def\HCv{E}
\begin{document}
\title{Generalized Noether theorem\\
in the realm of Hamiltonian point dynamics}

\author{J\"urgen Struckmeier}

\address{Frankfurt Institute for Advanced Studies (FIAS),\newline
Ruth-Moufang-Str.~1, 60438~Frankfurt am Main, Germany}

\ead{struckmeier@fias.uni-frankfurt.de}
\ifportablearticle
\date{}
\maketitle
\begin{center}
\small\paperaddress\par
\smallskip
\paperemail
\end{center}
\fi

\begin{abstract}
Noether's theorem in the realm of point dynamics establishes
the correlation of a constant of motion of a Hamilton-Lagrange system
with a particular symmetry transformation of its action functional.
Although usually derived in the Lagrangian formalism~\cite{noether,saletan},
the Hamiltonian formalism provides a natural context for this correlation.
Canonical transformations preserve the canonical form of the action up to a boundary term.
Within the class considered here, a symmetry additionally preserves the functional form of the specified extended Hamiltonian.
Every smooth phase-space first integral admits locally a corresponding
canonical Noether symmetry when momentum-dependent transformations
and boundary terms are allowed.
In order to represent symmetry mappings that include a transformation
of time, we use the extended Hamiltonian formalism.
This formalism enables us to define generating functions of canonical
transformations that also map time and energy in addition to the
conventional mappings of canonical space and momentum variables.
The physical energy hypersurface and the choice of the evolution parameter
are distinguished from identities on the full extended phase space.

As an example, a manifest representation of the symmetry transformation
associated with the Runge-Lenz invariant of the Kepler system is derived.
A suitable choice of evolution parameter makes its time-shifting
representative a symmetry of the specified extended action.
\end{abstract}

\section{Introduction}
Even more than a hundred years after the emergence of Einstein's
special theory of relativity, the presentation of classical
dynamics in terms of the Lagrangian and the Hamiltonian formalism
usually takes the physical time as the system evolution parameter.
This choice is compatible with relativistic as well as nonrelativistic
dynamics, but a more symmetric treatment is useful when time itself
is to be transformed.
It consists of introducing a system evolution parameter, $s$,
as the new independent variable, and of subsequently treating
the time $t=t(s)$ as a \emph{dependent} variable of $s$,
in parallel to all configuration space variables $q^{i}(s)$.

In order to preserve the canonical form of the action functional,
we introduce an \emph{extended} Hamiltonian, $H_{\e}$.
Setting up the correlation of extended and conventional Hamiltonians,
the \emph{crucial point} is that the energy coordinate $\Hv$
is varied independently of the conventional Hamiltonian $H$.
It agrees with the value of $H$ on the physical hypersurface $H-\Hv=0$.
Its negative, $-\Hv(s)$, plays the role of the
\emph{additional canonical variable} that is conjugate
to the time variable, $t(s)$.

With this relation of $H_{\e}$ and $H$ in place, the
extended set of canonical equations retains the conventional
canonical form.
A prescribed positive phase-space function specifies the relation
between $s$ and $t$ on physical trajectories.
The theory of extended canonical transformations retains its
canonical form as well.

It will be shown that the Hamiltonian form of Noether's theorem
can be represented by a one-parameter
\emph{infinitesimal canonical transformation}.
The characteristic function of the generator must be a constant
of motion throughout the extended phase-space domain in order
for the transformation to be a symmetry of the specified extended action.
Canonicity alone does not require this conservation condition.
The canonical transformation rules then embody the symmetry relations
that correspond to the constant of motion.
The statement is local and includes the boundary term of the action.
Hamiltonian Noether symmetries of this general kind and the role of
boundary conditions are also discussed in Ref.~\cite{mouchet}.

As a nontrivial example, we present a particular symmetry for the
Runge-Lenz invariant of the classical Kepler system that is associated
with a nonzero time shift.

\section{Generalized action functional}
The state of a classical dynamical system of $n$ degrees of freedom at time $t$
is completely described by $\bq=(q^{1},\ldots,q^{n})$ the vector
of generalized space coordinates and $\bp=(p_{1},\ldots,p_{n})$ the
covector of generalized momenta.
We assume the system to be described by a smooth Hamiltonian on the domain considered
\begin{equation}\label{H-conv-def}
H:\mathcal{U}\subset\mathbb{R}^{2n}\times\mathbb{R}\to\mathbb{R},\qquad
\Hv=H(\bq,\bp,t).
\end{equation}
A Hamiltonian $H$ contains the \emph{complete information}
on the given dynamical system through the dependence of its
\emph{value} $\Hv$
on each $q^{i}$ and each $p_{i}$ along the time axis $t$.

In order to formulate the principle of stationary action ---
originating from Leibniz, Maupertuis, Euler, and Lagrange ---
we define the action functional $\Phi(\gamma)$ as the line integral
\begin{equation}\label{actfunct0}
\Phi(\gamma)=\int\limits_{\gamma}
\sum_{i=1}^{n}p_{i}\dd q^{i}-H(\bq,\bp,t)\,\dd t,
\end{equation}
hence as a mapping of the set of phase-space paths
$\gamma\subset\mathbb{R}^{2n}\times\mathbb{R}$ into $\mathbb{R}$.
A phase-space path is a smooth curve
$\gamma(t)=(\bq(t),\bp(t),t)$ with $t_{0}\le t\le t_{1}$.
For the usual variational problem, the endpoint positions
$\bq(t_{0})=\bq_{0}$ and $\bq(t_{1})=\bq_{1}$ and the endpoint
times are fixed, while the momenta are varied independently.
The phase-space path $\gamma_{\mathrm{ext}}$ the dynamical system
actually realizes follows from the \emph{principle of stationary action}.
It states that the variation of the action functional~(\ref{actfunct0})
vanishes for $\gamma_{\mathrm{ext}}$, hence
$\delta\Phi(\gamma_{\mathrm{ext}})=0$.
Commonly, a restricted path $\gamma_{\mathrm{r}}\subset\RB^{2n}$ is defined
by parameterizing~(\ref{actfunct0}) in terms of the system's \emph{time}, $t$,
\begin{displaymath}
\gamma_{\mathrm{r}}(t)=\big(\bq(t),\bp(t)\big),\qquad t_{0}\le t\le t_{1}.
\end{displaymath}
The line integral~(\ref{actfunct0}) is thus converted into a
conventional integral.
The action principle then reads
\begin{equation}\label{actfunct1}
\delta\Phi(\gamma)=
\delta\int\limits_{t_{0}}^{t_{1}}\left[\sum_{i=1}^{n}p_{i}(t)
\dfrac{q^{i}(t)}{t}-H\big(\bq(t),\bp(t),t\big)\right]
\dd t\stackrel{!}{=}0.
\end{equation}
From the calculus of variations, one finds that the functional
$\Phi$ is \emph{stationary} at $\gamma_{\mathrm{ext}}$
($\delta\Phi(\gamma_{\mathrm{ext}})=0$) exactly if
the phase-space path $\big(\bq(t),\bp(t)\big)$
satisfies the ``canonical equations'' ($i=1,\dotsc,n$),
\begin{equation}\label{caneq-conv}
\dfrac{q^{i}}{t}=\pfrac{H}{p_{i}},\qquad
\dfrac{p_{i}}{t}=-\pfrac{H}{q^{i}},\qquad
\dfrac{\Hv}{t}=\pfrac{H}{t},
\end{equation}
where $\Hv(t)$ denotes according to Eq.~(\ref{H-conv-def})
the \emph{instantaneous value} of the Hamiltonian $H$.

The parametrization by $t$ remains valid when $H$ depends explicitly on time.
To treat transformations of time canonically, we instead let $t=t(s)$
and parameterize the line integral by an evolution parameter $s$~\cite{greiner},
\begin{displaymath}
\delta\int\limits_{s_{0}}^{s_{1}}\left[\sum_{i=1}^{n}
p_{i}(s)\,\dfrac{q^{i}(s)}{s}-H\big(
\bq(s),\bp(s),t(s)\big)\dfrac{t(s)}{s}\right]\dd s\stackrel{!}{=}0.
\end{displaymath}
We assume that $t(s)$ increases along the trajectories considered.
The $q^{i}$ and time $t$ are now treated on \emph{equal footing}.
The symmetric form of the integrand suggests introducing the
$2n+2$ dimensional \emph{extended phase space} with the additional pair
\begin{displaymath}
q^{0}(s)\equiv ct(s),\qquad p_{0}(s)\equiv -\Hv(s)/c.
\end{displaymath}
The constant $c$ supplies the conventional units and does not
restrict the construction to relativistic systems.
Here $\Hv$ is an independent energy coordinate.
On the physical hypersurface we impose
\begin{equation}\label{shellcondition}
\Hv=H(\bq,\bp,t)\qquad\Leftrightarrow\qquad H(\bq,\bp,t)-\Hv=0.
\end{equation}
Thus the distinction between the Hamiltonian function and its value
is implemented by extending the phase space.

We use the extended action in canonical form,
\begin{equation}\label{actfunct2}
\delta\Phi_{\e}=\delta\int\limits_{s_{0}}^{s_{1}}\left[
\sum_{\alpha=0}^{n}p_{\alpha}(s)\,\dfrac{q^{\alpha}(s)}{s}-
H_{\e}\big(\bq(s),\bp(s),t(s),\Hv(s)\big)\right]\dd s\stackrel{!}{=}0.
\end{equation}
On the physical hypersurface specified by Eq.~(\ref{shellcondition}), the canonical one-form
$p_{\alpha}\dd q^{\alpha}\equiv p_{i}\dd q^{i}-\Hv\dd t$ agrees with the integrand of the original action.
We therefore choose the extended Hamiltonian $H_{\e}$ to vanish on this hypersurface.
We require this zero to be regular, meaning that $\partial H_{\e}/\partial\Hv\ne0$ there,
so that the extended Hamiltonian flow advances physical time.

Since $\partial(H-\Hv)/\partial\Hv=-1$, every smooth extended
Hamiltonian that vanishes on this hypersurface admits locally
the factorization
\begin{equation}\label{H-ext-def}
H_{\e}(\bq,\bp,t,\Hv)=N(\bq,\bp,t,\Hv)\,C(\bq,\bp,t,\Hv),\qquad C(\bq,\bp,t,\Hv)=H(\bq,\bp,t)-\Hv.
\end{equation}
The regularity condition makes the smooth factor $N$ nonzero
on the physical hypersurface.
The canonical equation for $t$, derived below, gives
\begin{displaymath}
\left.\dfrac{t}{s}\right|_{\Hv=H}=-\left.\pfrac{H_{\e}}{\Hv}\right|_{\Hv=H}=N\big|_{\Hv=H}\;.
\end{displaymath}
The assumed increase of $t$ with $s$ therefore selects the
positive sign of $N$.
We work in a neighborhood where this factor remains positive.

Accordingly, we take $N$ to be a prescribed smooth positive
function without explicit dependence on $s$.
It is not varied as an independent multiplier.
It is dimensionless when $s$ has the units of time and specifies
the parametrization of physical trajectories in the extended
Hamiltonian formalism~\cite{struck1,struck2}.

The physical hypersurface is
\begin{equation}\label{H-ext-const}
\Sigma=\{(\bq,\bp,t,\Hv)\mid C=0\},\qquad
C\approx0,\qquad H_{\e}\approx0.
\end{equation}
The symbol $\approx$ denotes equality on $\Sigma$.
Neither $C$ nor $H_{\e}$ is identically zero on the full extended phase space.
Consequently, $H_{\e}$ must not be eliminated from the action before variation.

The action in Eq.~(\ref{actfunct2}) is varied over curves in the full extended phase space.
We vary $q^{i},p_{i},t,\Hv$ independently, with $q^{i}$ and $t$ fixed
at the endpoints and the momentum variations unrestricted.
The physical solutions are then selected by the initial condition
$C(s_{0})=0$.
The selected boundary data must be compatible with this condition.
The extended canonical equations are
\begin{equation}\label{caneq-ext}
\dfrac{q^{i}}{s}=\pfrac{H_{\e}}{p_{i}},\qquad
\dfrac{p_{i}}{s}=-\pfrac{H_{\e}}{q^{i}},\qquad
\dfrac{t}{s}=-\pfrac{H_{\e}}{\Hv},\qquad
\dfrac{\Hv}{s}=\pfrac{H_{\e}}{t}.
\end{equation}
For Eq.~(\ref{H-ext-def}), these equations read explicitly
\begin{align*}
\dfrac{q^{i}}{s}&=N\pfrac{H}{p_{i}}+C\pfrac{N}{p_{i}},&
\dfrac{p_{i}}{s}&=-N\pfrac{H}{q^{i}}-C\pfrac{N}{q^{i}},\\
\dfrac{t}{s}&=N\quad\;\;\,-C\pfrac{N}{\Hv},&
\dfrac{\Hv}{s}&=\hphantom{-}N\pfrac{H}{t}+C\pfrac{N}{t}.
\end{align*}
On $\Sigma$, we therefore have $\dd t/\dd s=N>0$.
Dividing the other equations by this expression recovers
the conventional canonical equations~(\ref{caneq-conv}).
In particular, $N$ agrees with $\dd t/\dd s$ on physical trajectories,
but these quantities must not generally be identified in an unrestricted
phase-space identity before differentiating $H_{\e}$.
The actions $\Phi_{\e}$ and $\Phi$ agree on physical trajectories.
Their equivalence as descriptions of physical motion follows from the
canonical equations and the initial condition $C(s_{0})=0$.

The extended phase space $T^{*}(\MB\times\RB)$, where $\MB$ is the
configuration manifold, is symplectic and has dimension $2n+2$.
Its symplectic form is
\begin{displaymath}
\omega_{\mathrm{ext}}=\sum_{i=1}^{n}\dd q^{i}\wedge\dd p_{i}
-\dd t\wedge\dd\Hv.
\end{displaymath}
The conventional phase space $T^{*}\MB$ is also symplectic.
The $2n+1$ dimensional evolution space $(T^{*}\MB)\times\RB$
and the physical hypersurface $\Sigma$ carry the associated
presymplectic description.

Since $H_{\e}$ has no explicit dependence on $s$, its total derivative is
\begin{align}
\dfrac{H_{\e}}{s}&=\sum_{i=1}^{n}\left(
\pfrac{H_{\e}}{p_{i}}\dfrac{p_{i}}{s}+
\pfrac{H_{\e}}{q^{i}}\dfrac{q^{i}}{s}\right)+
\pfrac{H_{\e}}{t}\dfrac{t}{s}+\pfrac{H_{\e}}{\Hv}\dfrac{\Hv}{s}\nonumber\\
&=0.\label{const-He}
\end{align}
Thus, if $\Hv(s_{0})=H(\bq(s_{0}),\bp(s_{0}),t(s_{0}))$, the condition
$H_{\e}=0$, and hence $C=0$, is maintained along the trajectory.
This is an invariant physical restriction on the initial data.
Its preservation does not remove the restriction, and the unconstrained
extended equations also admit solutions outside $\Sigma$.
The present fixed-parametrization formulation requires no additional
multiplier equation after the physical initial data have been specified.

For $N=1$, the relation $\dd t/\dd s=1$ holds identically,
so that the origins of $s$ and $t$ may be chosen to coincide.
If $H$ has no explicit time dependence, $\Hv$ is also constant along
physical trajectories.
The condition $H=\Hv_{0}$ then specifies their energy level.
Other conventional energy levels describe other admissible motions.
We will see in Sect.~\ref{sec:const-He} that conservation of $H_{\e}$
corresponds to the canonical flow generated by the extended equations.

\section{Example of a relativistic particle in an external potential $V$}
As an example for a \emph{non-trivial} extended Hamiltonian, we consider
$H_{\e}$ of a \emph{relativistic point particle} of mass $m>0$
in a prescribed potential energy $V(\bq,t)$ in the laboratory frame,
with no spatial vector potential,
\begin{equation}\label{H-ext-rp}
H_{\e}(\bq,\bp,t,\Hv)=\frac{1}{2m}
\left[\bp^{2}-{\left(\frac{\Hv-V(\bq,t)}{c}\right)}^{2}\right]+
\onehalf mc^{2}.
\end{equation}
On the physical hypersurface $H_{\e}=0$, Eq.~(\ref{H-ext-rp}) has
two energy branches.
We restrict to $\Hv-V>0$, which gives the equivalent
\emph{conventional} Hamiltonian $H$ as the right-hand side
of the equation $\Hv=H$,
\begin{equation}\label{H-conv-rp}
\Hv=\sqrt{\bp^{2}c^{2}+m^{2}c^{4}}+V(\bq,t)=H(\bq,\bp,t).
\end{equation}
To verify the relation to Eq.~(\ref{H-ext-def}) away from the
physical hypersurface, set
\begin{displaymath}
h(\bp)=\sqrt{\bp^{2}c^{2}+m^{2}c^{4}},\qquad
N=\frac{h(\bp)+\Hv-V}{2mc^{2}}.
\end{displaymath}
Then $H_{\e}=N(H-\Hv)$ is an exact phase-space identity.
The function $N$ is positive on the branch considered.
It equals $h/(mc^{2})$ on the physical hypersurface.

The canonical equation for $\dd t/\dd s$ is obtained as
\begin{displaymath}
\dfrac{t}{s}=-\pfrac{H_{\e}}{\Hv}=\frac{\Hv-V}{mc^{2}}=
\frac{\sqrt{\bp^{2}c^{2}+m^{2}c^{4}}}{mc^{2}}=
\sqrt{1+{\left(\frac{\bp}{mc}\right)}^{2}}=\gamma.
\end{displaymath}
The equalities involving the square root hold on the positive-energy
physical hypersurface.
The remaining equations are
\begin{displaymath}
\dfrac{\bq}{s}=\frac{\bp}{m},\qquad
\dfrac{\bp}{s}=-\frac{\Hv-V}{mc^{2}}\,\boldsymbol{\nabla}V,\qquad
\dfrac{\Hv}{s}=\frac{\Hv-V}{mc^{2}}\pfrac{V}{t}.
\end{displaymath}
Dividing these equations by $\dd t/\dd s$ gives the conventional
canonical equations generated by $H$ on that hypersurface.
Moreover, the velocity relation gives
$\gamma=(1-|\dd\bq/\dd t|^{2}/c^{2})^{-1/2}$.
Thus, if $t$ is laboratory time, $s$ measures the particle's
\emph{proper time} on physical trajectories.
Thus, $H_{\e}$ from Eq.~(\ref{H-ext-rp}) and $H$ from Eq.~(\ref{H-conv-rp})
indeed describe \emph{the same physical system}.

Setting up the canonical equations for $\dd\bq/\dd s$,
\begin{displaymath}
\dfrac{\bq}{s}=\pfrac{H_{\e}}{\bp}=\frac{\bp}{m},
\end{displaymath}
we derive the equivalent \emph{extended Lagrangian} $L_{\e}$
by means of the Legendre transformation
\begin{displaymath}
L_{\e}\left(\bq,\dfrac{\bq}{s},t,\dfrac{t}{s}\right)=
\bp\,\dfrac{\bq}{s}-\Hv\,\dfrac{t}{s}-H_{\e}\big(\bq,\bp,t,\Hv\big),
\end{displaymath}
which yields
\begin{displaymath}
L_{\e}=\onehalf mc^{2}\left[\frac{1}{c^{2}}{\left(\dfrac{\bq}{s}\right)}^{2}-
{\left(\dfrac{t}{s}\right)}^{2}-1\right]-V(\bq,t)\dfrac{t}{s}.
\end{displaymath}
On $H_{\e}=0$, the physical normalization is
\begin{displaymath}
{\left(\dfrac{t}{s}\right)}^{2}=1+\frac{1}{c^{2}}{\left(\dfrac{\bq}{s}\right)}^{2},\qquad
\dfrac{s}{t}=\gamma^{-1}=\sqrt{1-\frac{1}{c^{2}}{\left(\dfrac{\bq}{t}\right)}^{2}},
\end{displaymath}
and the well-known conventional Lagrangian $L$ is recovered on
physically normalized trajectories via
\begin{align*}
L\left(\bq,\dfrac{\bq}{t},t\right)&=L_{\e}\dfrac{s}{t}=-mc^{2}\dfrac{s}{t}-V(\bq,t)\\
&=-mc^{2}\sqrt{1-\frac{1}{c^{2}}{\left(\dfrac{\bq}{t}\right)}^{2}}-V(\bq,t).
\end{align*}
Similar to the conventional Hamiltonian $H$ from Eq.~(\ref{H-conv-rp}),
the conventional Lagrangian $L$ of the relativistic point particle is no
longer a quadratic form in the velocities.
The physical normalization is used after deriving the extended
equations of motion and must not be substituted into $L_{\e}$
before its unrestricted variation.

The conventional Hamiltonian $H$ is no longer a quadratic form in the momenta.
The quadratic extended constraint is thus particularly convenient for
obtaining a local wave equation by formal canonical quantization.
With
\begin{displaymath}
p_{\mu}\mapsto\hat{p}_{\mu}=-i\hbar\pfrac{}{q^{\mu}},
\qquad\Hv\mapsto\hat{\Hv}=i\hbar\pfrac{}{t},
\end{displaymath}
the condition $\hat H_{\e}\psi=0$ yields
\begin{displaymath}
\left[\left(i\hbar\pfrac{}{t}-V\right)^{2}
+\hbar^{2}c^{2}\nabla^{2}-m^{2}c^{4}\right]\psi=0.
\end{displaymath}
This is the Klein-Gordon equation for the stated potential coupling.
The square denotes the full operator product,
including derivatives of $V$ when it depends on time.
On the other hand, the square-root Hamiltonian can also be quantized as a spatially nonlocal operator.
Its quantum theory is not automatically equivalent to this constraint
quantization for a general $V$, since $V$ need not commute with the kinetic operator.
The nonquadratic form of the conventional functions does not
prevent canonical quantization or a phase-space path integral.
It does mean that the corresponding momentum integrations need not be Gaussian.
\section{Extended canonical transformations}
Canonical transformations preserve the canonical form of the action up to a boundary term.
Within the class considered here, a symmetry additionally preserves the functional form of the specified extended Hamiltonian.
In the extended description, time $q^{0}\equiv ct$
and $p_{0}\equiv-\Hv/c$ are a conjugate pair of dynamical variables.
The local defining relation is
\begin{displaymath}
\sum_{i=1}^{n}p_{i}\dd q^{i}-\Hv\dd t-H_{\e}\dd s
=\sum_{i=1}^{n}P_{i}\dd Q^{i}-\HCv\dd T-H_{\e}^{\,\prime}\dd s
+\dd F_{1}.
\end{displaymath}
The boundary term and the endpoint conditions must be transformed
consistently in the variational principle.
For the local equations, one may equivalently use variations supported
away from the boundaries.
We first consider generators with no explicit dependence on $s$.
Whenever a generating function of type $F_{1}(\bq,\bQ,t,T)$ exists, its total differential is
\begin{displaymath}
\dd F_1=\sum_{i=1}^{n}\pfrac{F_{1}}{q^{i}}\dd q^i+\sum_{i=1}^{n}\pfrac{F_{1}}{Q^{i}}\dd Q^i
+\pfrac{F_{1}}{t}\dd t+\pfrac{F_{1}}{T}\dd T.
\end{displaymath}
Comparison of coefficients gives
\begin{displaymath}
p_{i}=\hphantom{-}\pfrac{F_{1}}{q^{i}},\qquad
P_{i}=-\pfrac{F_{1}}{Q^{i}},\qquad
\Hv=-\pfrac{F_{1}}{t},\qquad
\HCv=\hphantom{-}\pfrac{F_{1}}{T},\qquad
H_{\e}^{\,\prime}=H_{\e}.
\end{displaymath}
Here the last equality means
$H_{\e}^{\,\prime}(\bQ,\bP,T,\HCv)=H_{\e}(\bq,\bp,t,\Hv)$
at corresponding points.
It preserves the zero value on physical trajectories.
It does not require the same functional dependence of $H_{\e}^{\,\prime}$
and $H_{\e}$ on their respective arguments.

The Legendre transformation
\begin{displaymath}
F_{2}(\bq,\bP,t,\HCv)=F_{1}(\bq,\bQ,t,T)+\sum_{i=1}^{n}Q^{i}P_{i}-T\HCv
\end{displaymath}
yields an equivalent set of canonical transformation rules
where this change of generating variables is nonsingular,
\begin{equation}\label{rules-F2}
p_{i}=\hphantom{-}\pfrac{F_{2}}{q^{i}},\qquad
Q^{i}=\hphantom{-}\pfrac{F_{2}}{P_{i}},\qquad
\Hv=-\pfrac{F_{2}}{t},\qquad T=-\pfrac{F_{2}}{\HCv},\qquad
H_{\e}^{\,\prime}=H_{\e}.
\end{equation}
A type-$F_{2}$ representation exists locally for transformations
sufficiently close to the identity.
For generators with explicit $s$ dependence, the last rule is replaced by
$H_{\e}^{\,\prime}=H_{\e}+\partial F_{2}/\partial s$.

We now return to generating functions without explicit dependence on $s$.
To recover a conventional transformed Hamiltonian, we restrict to a
branch on which the transformed energy hypersurface can be solved as
$\HCv=H^{\,\prime}(\bQ,\bP,T)$.
Locally this is possible if $\partial H_{\e}^{\,\prime}/\partial\HCv\ne0$
on that hypersurface.
To preserve the time orientation, we require
$-\partial H_{\e}^{\,\prime}/\partial\HCv>0$ there.
The transformed extended Hamiltonian then admits a factorization
$H_{\e}^{\,\prime}=N^{\,\prime}(H^{\,\prime}-\HCv)$
with $N^{\,\prime}>0$ in a suitable neighborhood.
According to Eq.~(\ref{H-ext-def}), the exact transformation rule is
\begin{displaymath}
N^{\,\prime}\Big[H^{\,\prime}(\bQ,\bP,T)-\HCv\Big]
=N\Big[H(\bq,\bp,t)-\Hv\Big].
\end{displaymath}
Equivalently,
\begin{equation}\label{rules-conv-H}\begin{split}
\Big[H^{\,\prime}(\bQ,\bP,T)-\HCv\Big]\frac{N^{\,\prime}}{N}
&=H(\bq,\bp,t)-\Hv,\\
\Big[H(\bq,\bp,t)-\Hv\Big]\frac{N}{N^{\,\prime}}
&=H^{\,\prime}(\bQ,\bP,T)-\HCv.
\end{split}
\end{equation}
On physical trajectories,
\begin{displaymath}
\left.\frac{N^{\,\prime}}{N}\right|_{\Sigma}
=\frac{\dd T/\dd s}{\dd t/\dd s}=\dfrac{T}{t}.
\end{displaymath}
This identification with the time-scale ratio holds on the physical
hypersurface and must not replace the phase-space prefactors
in Eq.~(\ref{rules-conv-H}) before differentiation.

\subsection{Extended generating function of a
conventional canonical transformation}
An important example of an extended generating function
is the particular $F_{2}$ that defines a \emph{conventional}
canonical transformation.
Consider the particular extended generating function
\begin{equation}\label{gen-conv}
F_{2}(\bq,\bP,t,\HCv)=f_{2}(\bq,\bP,t)-t\HCv,
\end{equation}
with $f_{2}(\bq,\bP,t)$ denoting a conventional
generating function.
The coordinate transformation rules~(\ref{rules-F2})
for this $F_{2}$ follow as
\begin{displaymath}
p_{i}=\pfrac{f_{2}}{q^{i}},\qquad Q^{i}=\pfrac{f_{2}}{P_{i}},
\qquad\Hv=-\pfrac{f_{2}}{t}+\HCv,\qquad T=t.
\end{displaymath}
Choose $N^{\,\prime}$ as $N$ expressed in the new canonical
variables, so that their values at corresponding points agree.
Since $T=t$, Eq.~(\ref{rules-conv-H}) then gives
\begin{displaymath}
H^{\,\prime}-\HCv=H-\Hv\qquad\Longrightarrow\qquad
H^{\,\prime}(\bQ,\bP,t)=H(\bq,\bp,t)+\pfrac{f_{2}}{t}.
\end{displaymath}
The partial derivatives of $f_{2}$ obviously yield
the usual conventional canonical transformation rules.
The particular extended generating function $F_{2}$ from
Eq.~(\ref{gen-conv}) thus defines the conventional
canonical transformation generated by $f_{2}$.
We conclude that the group of conventional canonical
transformations establishes a \emph{subgroup} of the
group of extended canonical transformations.
\section{Generalized Noether theorem}
We are now prepared to derive the generalized Noether theorem in the
Hamiltonian formalism on the basis of an extended infinitesimal
canonical transformation.
The type-$F_{2}$ generating function of such a transformation has the form
\begin{equation}\label{gen-infini}
F_{2}(\bq,\bP,t,\HCv)=\sum_{i=1}^{n}q^{i}P_{i}-t\HCv+
\epsilon\,I(\bq,\bP,t,\HCv)+\mathcal O(\epsilon^{2}),
\end{equation}
where $\epsilon$ is a small constant parameter and $I$ is a smooth
function with no explicit dependence on $s$.
The characteristic term must be written in the mixed variables
appropriate to $F_{2}$.
Only after expanding to first order may its derivatives be evaluated
at the original variables $(\bq,\bp,t,\Hv)$.
All infinitesimal transformation equations below are understood
through first order in $\epsilon$.
The transformation rules~(\ref{rules-F2}) give
\begin{align*}
p_{i}&=P_{i}+\epsilon\pfrac{I}{q^{i}},&
\Hv&=\HCv-\epsilon\pfrac{I}{t},\\
Q^{i}&=q^{i}+\epsilon\pfrac{I}{P_{i}},&
T&=t-\epsilon\pfrac{I}{\HCv},\qquad H_{\e}^{\,\prime}=H_{\e}.
\end{align*}
Thus the variations of the original canonical variables are
\begin{equation}\label{rules-infini}
\begin{aligned}
\delta p_{i}&=-\epsilon\pfrac{I}{q^{i}},&
\delta\Hv&=\hphantom{-}\epsilon\pfrac{I}{t},\\
\delta q^{i}&=\hphantom{-}\epsilon\pfrac{I}{p_{i}},&
\delta t&=-\epsilon\pfrac{I}{\Hv}.
\end{aligned}
\end{equation}
Every smooth $I$ defines a local canonical flow.
The transformation law
$H_{\e}^{\,\prime}(\bQ,\bP,T,\HCv)=H_{\e}(\bq,\bp,t,\Hv)$
holds for the appropriately transformed Hamiltonian.
A symmetry of the specified extended system additionally requires
$H_{\e}^{\,\prime}$ to be the same function as $H_{\e}$.
It therefore requires
$H_{\e}(\bQ,\bP,T,\HCv)=H_{\e}(\bq,\bp,t,\Hv)$.

The first-order change of this fixed function is
\begin{align}
\delta H_{\e}&=\epsilon\left[\sum_{i=1}^{n}
\left(\pfrac{H_{\e}}{q^{i}}\pfrac{I}{p_{i}}-
\pfrac{H_{\e}}{p_{i}}\pfrac{I}{q^{i}}\right)-
\pfrac{H_{\e}}{t}\pfrac{I}{\Hv}+
\pfrac{H_{\e}}{\Hv}\pfrac{I}{t}\right]\nonumber\\
&=\epsilon\,{\left[H_{\e},I\right]}_{\mathrm{ext}}.
\label{commute-cond}
\end{align}
For any two phase-space functions $f,g$, the extended Poisson bracket is
\begin{displaymath}
{\left[f,g\right]}_{\mathrm{ext}}
=\sum_{i=1}^{n}\left(\pfrac{f}{q^{i}}\pfrac{g}{p_{i}}
-\pfrac{f}{p_{i}}\pfrac{g}{q^{i}}\right)
-\pfrac{f}{t}\pfrac{g}{\Hv}+\pfrac{f}{\Hv}\pfrac{g}{t}.
\end{displaymath}
Along any solution of the extended canonical equations,
\begin{displaymath}
\dfrac{I}{s}={\left[I,H_{\e}\right]}_{\mathrm{ext}},\qquad
\delta H_{\e}=-\epsilon\dfrac{I}{s}.
\end{displaymath}
Thus the symmetry condition is equivalent to conservation of $I$
along every extended trajectory in the domain considered.
For this statement, the bracket must vanish as a phase-space identity,
and not merely on the physical hypersurface $\Sigma$.

\subsection{Derivation of the boundary function}\label{sec:boundary-function}
The boundary term can be found directly without using the equations of motion.
We write the action in Eq.~(\ref{actfunct2}) as
\begin{displaymath}
\Phi_{\e}=\int_{s_{0}}^{s_{1}}\mathcal{L}_{\mathrm{can}}\,\dd s,\qquad
\mathcal{L}_{\mathrm{can}}=
\sum_{i=1}^{n}p_{i}\dfrac{q^{i}}{s}-\Hv\dfrac{t}{s}-H_{\e}.
\end{displaymath}
The infinitesimal transformation~(\ref{rules-infini}) acts at fixed $s$ with fixed integration limits and constant $\epsilon$.
Thus $\delta$ commutes with $\dd/\dd s$.
Varying the action and integrating the terms $p_{i}\,\dd(\delta q^{i})/\dd s$ and $-\Hv\,\dd(\delta t)/\dd s$ by parts gives
\begin{align*}
\delta\Phi_{\e}={}&
\left[\sum_{i=1}^{n}p_{i}\delta q^{i}-\Hv\delta t\right]_{s_{0}}^{s_{1}}\\
&+\int_{s_{0}}^{s_{1}}\left[
\sum_{i=1}^{n}\left(\dfrac{q^{i}}{s}\delta p_{i}-\dfrac{p_{i}}{s}\delta q^{i}\right)
-\dfrac{t}{s}\delta\Hv+\dfrac{\Hv}{s}\delta t-\delta H_{\e}
\right]\dd s.
\end{align*}
Substituting the canonical variations from Eq.~(\ref{rules-infini}) into the kinetic terms under the integral yields
\begin{align*}
&\sum_{i=1}^{n}\left(\dfrac{q^{i}}{s}\delta p_{i}-\dfrac{p_{i}}{s}\delta q^{i}\right)
-\dfrac{t}{s}\delta\Hv+\dfrac{\Hv}{s}\delta t\\
&\qquad=-\epsilon\left[
\sum_{i=1}^{n}\left(\pfrac{I}{q^{i}}\dfrac{q^{i}}{s}
+\pfrac{I}{p_{i}}\dfrac{p_{i}}{s}\right)
+\pfrac{I}{t}\dfrac{t}{s}+\pfrac{I}{\Hv}\dfrac{\Hv}{s}
\right]\\
&\qquad=-\epsilon\dfrac{I}{s}.
\end{align*}
The last equality is the chain rule for $I(\bq,\bp,t,\Hv)$, which has no explicit dependence on $s$.
This identity holds along any smooth curve in the extended phase space, without imposing the equations of motion or the constraint $\Hv=H$.
Consequently,
\begin{displaymath}
\delta\Phi_{\e}
=\left[\sum_{i=1}^{n}p_{i}\delta q^{i}-\Hv\delta t-\epsilon I\right]_{s_{0}}^{s_{1}}
-\int_{s_{0}}^{s_{1}}\delta H_{\e}\,\dd s.
\end{displaymath}
Using $\delta q^{i}=\epsilon\,\partial I/\partial p_{i}$ and $\delta t=-\epsilon\,\partial I/\partial\Hv$, we identify the boundary function
\begin{displaymath}
B=\sum_{i=1}^{n}p_{i}\pfrac{I}{p_{i}}+\Hv\pfrac{I}{\Hv}-I.
\end{displaymath}
Together with Eq.~(\ref{commute-cond}), this gives
\begin{align*}
\delta\mathcal{L}_{\mathrm{can}}
&=\epsilon\dfrac{B}{s}
+\epsilon{\left[I,H_{\e}\right]}_{\mathrm{ext}},\\
\delta\Phi_{\e}
&=\epsilon\big[B\big]_{s_{0}}^{s_{1}}
+\epsilon\int_{s_{0}}^{s_{1}}{\left[I,H_{\e}\right]}_{\mathrm{ext}}\,\dd s.
\end{align*}
For a generator that commutes identically with $H_{\e}$, the action variation reduces to
\begin{displaymath}
\delta\Phi_{\e}=\epsilon\big[B\big]_{s_{0}}^{s_{1}}.
\end{displaymath}
The action is therefore invariant up to this boundary term.
The term is generally nonzero, and the boundary conditions must be
transformed consistently with the canonical map.

\subsection{Theorem and its local inverse}
We can now state the generalized Noether theorem and its local inverse.
\newtheorem{theorem}{Theorem}
\begin{theorem}[generalized Noether]
Let $H_{\e}$ be a smooth extended Hamiltonian without an explicit dependence on $s$.
If a smooth phase-space function $I$ satisfies
${\left[I,H_{\e}\right]}_{\mathrm{ext}}=0$ identically, then
Eq.~(\ref{rules-infini}) generates a local one-parameter canonical
symmetry of the extended action up to a boundary term.
The function $I$ is conserved along every extended Hamiltonian trajectory.

Conversely, every smooth infinitesimal variational symmetry at fixed $s$
whose transformation coefficients and boundary function depend only on
the extended phase-space coordinates admits locally such a generator $I$.
In particular, neither has explicit dependence on $s$.
It is determined by that symmetry up to an additive constant.
\end{theorem}
For the converse, define the canonical $1$-form
\begin{displaymath}
\theta=\sum_{i=1}^{n}p_{i}\dd q^{i}-\Hv\dd t,\qquad\dd\theta=-\omega_{\mathrm{ext}}
\end{displaymath}
and denote the infinitesimal symmetry vector field by $X$.
Comparison of the velocity coefficients in $\delta\mathcal{L}_{\mathrm{can}}=\epsilon\,\dd B/\dd s$ gives
\begin{displaymath}
\mathcal{L}_{X}\theta=\dd B.
\end{displaymath}
Cartan's identity then yields
\begin{displaymath}
\iota_{X}\dd\theta=-\dd I,\quad\text{with}\quad I=\theta(X)-B.
\end{displaymath}
Here $\mathcal{L}_{X}$ denotes the Lie derivative along $X$ and $\iota_{X}$ denotes contraction with $X$.
The vector field $X$ is normalized so that $\delta f=\epsilon X(f)$ for every phase-space function $f$.
Since $\omega_{\mathrm{ext}}$ is nondegenerate, $X$ has precisely the canonical form in Eq.~(\ref{rules-infini}).
The remaining coefficient gives
\begin{displaymath}
{\left[I,H_{\e}\right]}_{\mathrm{ext}}=0.
\end{displaymath}
If explicit $s$ dependence is allowed, the corresponding conservation
condition is
\begin{displaymath}
\pfrac{I}{s}+{\left[I,H_{\e}\right]}_{\mathrm{ext}}=0,
\end{displaymath}
where the partial derivative holds all canonical variables fixed.

We may reformulate the physical conservation law in terms of the
conventional Hamiltonian $H$ with time $t$ as the independent variable.
For $H_{\e}=NC$, the exact bracket identity is
\begin{displaymath}
{\left[I,H_{\e}\right]}_{\mathrm{ext}}
=N{\left[I,C\right]}_{\mathrm{ext}}
+C{\left[I,N\right]}_{\mathrm{ext}},
\end{displaymath}
where
\begin{displaymath}
{\left[I,C\right]}_{\mathrm{ext}}=
\pfrac{I}{t}+\pfrac{I}{\Hv}\pfrac{H}{t}
+\sum_{i=1}^{n}\left(\pfrac{I}{q^{i}}\pfrac{H}{p_{i}}
-\pfrac{I}{p_{i}}\pfrac{H}{q^{i}}\right).
\end{displaymath}
On physical trajectories, $C=0$ and $\dd t/\dd s=N$.
The strong extended conservation condition therefore implies
\begin{equation}\label{invar-t}
\dfrac{}{t}I\big(\bq(t),\bp(t),t,\Hv(t)\big)=0,\qquad \Hv(t)=H(\bq(t),\bp(t),t).
\end{equation}
The relation $\Hv=H$ specifies the physical hypersurface on which $I$ is evaluated.
The associated canonical transformation is generated by the full function $I(\bq,\bp,t,\Hv)$.
Its partial derivatives are taken with all other extended canonical coordinates fixed.
For physical trajectories, the condition $\Hv=H$ is imposed only after taking these derivatives.
The infinitesimal transformation rules are
\begin{equation}\label{rules-infini-t}
\delta p_{i}=-\epsilon\pfrac{I}{q^{i}},\qquad
\delta q^{i}=\epsilon\pfrac{I}{p_{i}},\qquad
\delta\Hv=\epsilon\pfrac{I}{t},\qquad
\delta t\big|_\Sigma=\left.-\epsilon\pfrac{I}{\Hv}\right|_{\Hv=H}.
\end{equation}
Conservation only on $\Sigma$ does not in general establish a
symmetry of a fixed extended action for an arbitrary $N$.
The choice of the extension $I$ and of $N$ matters.

Nevertheless, every smooth conventional first integral has a local Noether representation.
Indeed, if $I(\bq,\bp,t)$ satisfies $\partial I/\partial t+[I,H]_{\bq,\bp}=0$ throughout its domain,
choose its extension independent of $\Hv$ and set $N=1$.
Then $[I,H-\Hv]_{\mathrm{ext}}=0$ identically, so the theorem applies.
This statement allows momentum-dependent symmetry transformations.
It does not assert that every invariant arises from a point
transformation of configuration space or that the symmetry flow
exists globally.

\section{Examples}
\subsection{Symmetry transformation associated with constant extended Hamiltonian $H_{\e}$}\label{sec:const-He}
From Eq.~(\ref{const-He}) we know that $H_{\e}$ is constant along
every extended trajectory.
Its value is zero on physical trajectories.
According to the generalized Noether theorem, the generating function
\begin{displaymath}
F_{2}(\bq,\bP,t,\HCv)=\sum_{i=1}^{n}q^{i}P_{i}-t\HCv+
\delta s\,H_{\e}(\bq,\bP,t,\HCv)+\mathcal O(\delta s^{2})
\end{displaymath}
then defines the corresponding one-parameter symmetry transformation,
which is established by the set of canonical transformation rules.
The general set of transformation rules from Eqs.~(\ref{rules-infini})
reads for the above generating function $F_{2}$
\begin{align*}
P_{i}&=p_{i}-\delta s\pfrac{H_{\e}}{q^{i}},&
\HCv&=\Hv+\delta s\pfrac{H_{\e}}{t}\\
Q^{i}&=q^{i}+\delta s\pfrac{H_{\e}}{p_{i}},&
T&=t-\delta s\pfrac{H_{\e}}{\Hv}\\
H_{\e}^{\,\prime}&=H_{\e}.
\end{align*}
Inserting the extended set of canonical equations from Eqs.~(\ref{caneq-ext}),
the symmetry transformation rules for the canonical coordinates are converted into
\begin{align*}
P_{i}&=p_{i}+\dfrac{p_{i}}{s}\delta s,&
\HCv&=\Hv+\dfrac{\Hv}{s}\delta s\\
Q^{i}&=q^{i}+\dfrac{q^{i}}{s}\delta s,&
T&=t+\dfrac{t}{s}\delta s.
\end{align*}
The system is thus shifted an infinitesimal step $\delta s$
along the system's evolution parameter, $s$.
Integrating this Hamiltonian vector field gives the corresponding
finite canonical flow wherever the solution exists.
The first-order formulas themselves are not exact finite-step maps.
Thus the symmetry associated with conservation of $H_{\e}$ is the Hamiltonian evolution in $s$,
which preserves the symplectic structure and the extended action up to a boundary term.
On the physical hypersurface, this transformation moves a point
along its physical trajectory.

\subsection{Rotational symmetry and angular momentum conservation of a Kepler system}
The classical Kepler system is a two-body problem with
the mutual interaction following an inverse square force law.
For constant masses $m_{1}$ and $m_{2}$, let $q_{1},q_{2}$ denote
the Cartesian relative coordinates in the plane of motion.
In these Euclidean coordinates we write spatial indices as subscripts.
We use the Hamiltonian and action per unit reduced mass
$m_{\mathrm r}=m_{1}m_{2}/(m_{1}+m_{2})$.
The corresponding canonical momenta are $p_{i}=\dd q_{i}/\dd t$,
and the physical relative momenta are $m_{\mathrm r}p_{i}$.
The relative equations of motion are
\begin{equation}\label{kepler}
\frac{\dd^{2}q_{i}}{\dd t^{2}}+\mu(t)\frac{q_{i}}
{{\big(q_{1}^{2}+q_{2}^{2}\big)}^{3/2}}=0,\qquad i=1,2.
\end{equation}
For fixed gravitational masses, $\mu=G(m_{1}+m_{2})$ is constant.
To include an explicitly time-dependent central field, we also allow
$\mu(t)$ to be a prescribed smooth coupling.
This extension does not describe arbitrary time-dependent
masses or mass-transfer dynamics.

The equations of motion~(\ref{kepler}) follow from the Hamiltonian
\begin{equation}\label{kepham}
H(\bq,\bp,t)=\onehalf p_{1}^{2}+\onehalf p_{2}^{2}+V(\bq,t)
\end{equation}
containing the interaction potential
\begin{displaymath}
V(\bq,t)=-\frac{\mu(t)}{\sqrt{q_{1}^{2}+q_{2}^{2}}}=-\frac{\mu(t)}{r}.
\end{displaymath}
As the potential spatially depends on the distance
$r=\sqrt{q_{1}^{2}+q_{2}^{2}}$ only, it is obviously
invariant with respect to rotations in
configuration space, $(q_{1},q_{2})$
\begin{equation}\label{Kepler-rotation}
\begin{pmatrix}Q_{1}\\ Q_{2}\end{pmatrix}=
\begin{pmatrix}\hphantom{-}\cos\epsilon&\sin\epsilon\\
-\sin\epsilon&\cos\epsilon\end{pmatrix}
\begin{pmatrix}q_{1}\\ q_{2}\end{pmatrix},
\end{equation}
with positive $\epsilon$ denoting a clockwise active rotation.
The angle is constant, and the rotation matrix is independent of
the phase-space point.
This is a rigid one-parameter rotation.

This symmetry is maintained if we choose $|\epsilon|\ll1$.
Then $\cos\epsilon\approx 1$, $\sin\epsilon\approx\epsilon$,
and the infinitesimal rules are
\begin{displaymath}
\delta q_{1}\equiv Q_{1}-q_{1}=\epsilon\,q_{2},\qquad
\delta q_{2}\equiv Q_{2}-q_{2}=-\epsilon\,q_{1}.
\end{displaymath}
These rules can be regarded as being derived from the generating
function of the \emph{infinitesimal} canonical transformation
\begin{equation}\label{Kepler-rotation-gen}
F_{2}(q_{1},q_{2},P_{1},P_{2},t,\HCv)=-t\HCv+q_{1}P_{1}+q_{2}P_{2}+
\epsilon\,(P_{1}q_{2}-P_{2}q_{1})+\mathcal O(\epsilon^{2}).
\end{equation}
Both the coordinates and the momenta undergo the same rotation.
Choose $N=1$, so that $H_{\e}=H-\Hv$.
The rotational invariance of this Hamiltonian implies that the
generator commutes with $H_{\e}$ and is conserved.
Thus
\begin{displaymath}
I=p_{1}q_{2}-p_{2}q_{1}=-L_{z},\qquad\dfrac{I}{t}=0,
\end{displaymath}
where $L_{z}=q_{1}p_{2}-q_{2}p_{1}$ is the angular momentum per unit
reduced mass.
This establishes the well-known \emph{conservation law of angular
momentum} in possibly time-dependent central-force fields.
With a given symmetry transformation, we have thus determined
the corresponding constant of motion.

As with any generating function of a canonical transformation,
we can derive from the generating function~(\ref{Kepler-rotation-gen})
the rules of both the configuration space coordinates and the
respective canonical momenta.
In matrix form, the infinitesimal rules for the momenta are
\begin{displaymath}
\begin{pmatrix}P_{1}\\ P_{2}\end{pmatrix}=
\left[\Eins+\AB_{\epsilon}\right]
\begin{pmatrix}p_{1}\\ p_{2}\end{pmatrix},\qquad
\AB_{\epsilon}=\epsilon
\begin{pmatrix}\hphantom{-}0&1\\-1&0\end{pmatrix},
\end{displaymath}
with $\Eins$ denoting the $2\times 2$ unit matrix.
The corresponding \emph{finite} transformation is then
\begin{displaymath}
\begin{pmatrix}P_{1}\\ P_{2}\end{pmatrix}=
\exp{(\AB_{\epsilon})}
\begin{pmatrix}p_{1}\\ p_{2}\end{pmatrix},\qquad
\exp{(\AB_{\epsilon})}=\begin{pmatrix}\hphantom{-}\cos\epsilon&
\sin\epsilon\\ -\sin\epsilon&\cos\epsilon\end{pmatrix},
\end{displaymath}
which coincides with the transformation of the configuration
space variables from Eq.~(\ref{Kepler-rotation}).
This reflects the fact that the Hamiltonian~(\ref{kepham})
is equally invariant under rotations in momentum space.
\subsection{Symmetry transformations associated
with the Runge-Lenz invariant of the Kepler system}
Conversely, if $I$ denotes a smooth first integral of a Hamiltonian
system, its local canonical symmetry is obtained from
the generating function~(\ref{gen-infini}) with the qualifications
stated in the theorem.
As an example, a particularly transparent representation of the
symmetry associated with the Runge-Lenz invariant of the
time-independent Kepler system~(\ref{kepham}) is derived by admitting
a transformation $(q_{1},q_{2})|_{t}\mapsto(Q_{1},Q_{2})|_{t+\delta t}$
with a \emph{time shift} $\delta t$.
Throughout this example, $\mu$ is constant and $r>0$.

One component of the constant Runge-Lenz vector, with an overall minus
sign relative to the usual first component, is
\begin{equation}\label{rl-1}
I_{0}(q_{1},q_{2},p_{1},p_{2})=-q_{1}p_{2}^{2}+q_{2}p_{1}p_{2}+
\mu\frac{q_{1}}{\sqrt{q_{1}^{2}+q_{2}^{2}}}.
\end{equation}
For $N=1$, this function satisfies
$[I_{0},H-\Hv]_{\mathrm{ext}}=0$ identically.
Using it in Eq.~(\ref{gen-infini}) therefore gives a symmetry of
the extended action with $H_{\e}=H-\Hv$.
As $I_{0}$ is independent of $\Hv$, its transformation has $\delta t=0$.

A more conspicuous representation emerges if the invariant is
extended differently away from the physical hypersurface.
On $\Sigma$ we have
\begin{displaymath}
\Hv=\onehalf p_{1}^{2}+\onehalf p_{2}^{2}
-\frac{\mu}{\sqrt{q_{1}^{2}+q_{2}^{2}}}.
\end{displaymath}
Replacing the square-root term in Eq.~(\ref{rl-1}) on this
hypersurface gives the representative
\begin{equation}\label{rl-1a}
I(q_{1},q_{2},p_{1},p_{2},\Hv)=\onehalf q_{1}p_{1}^{2}+q_{2}p_{1}p_{2}
-\onehalf q_{1}p_{2}^{2}-q_{1}\Hv.
\end{equation}
These functions agree on physical trajectories, but their extensions
to the full extended phase space are different.
Their exact relation is
\begin{displaymath}
I=I_{0}+q_{1}C,\qquad C=H-\Hv.
\end{displaymath}
In particular,
\begin{displaymath}
[I,C]_{\mathrm{ext}}=p_{1}C,\qquad
[I,r]_{\mathrm{ext}}=-p_{1}r.
\end{displaymath}
Thus $I$ is only weakly conserved for the unit choice $N=1$.
To obtain a symmetry of a fixed extended action, we now choose the
parametrization function
\begin{displaymath}
N=\frac{r}{\ell},\qquad
H_{\e}=\frac{r}{\ell}(H-\Hv),\qquad
\left.\dfrac{t}{s}\right|_{\Sigma}=\frac{r}{\ell},
\end{displaymath}
where $\ell>0$ is a fixed reference length.
This changes the parametrization, while the conventional Kepler
equations in physical time remain unchanged.
The two bracket terms cancel identically,
\begin{displaymath}
[I,H_{\e}]_{\mathrm{ext}}
=\frac{r}{\ell}\,p_{1}C-\frac{p_{1}r}{\ell}\,C=0.
\end{displaymath}
The representative~(\ref{rl-1a}) therefore meets the strong symmetry
condition of the theorem for this choice of $H_{\e}$.

The dependence on the canonical energy variable $\Hv$ now gives
a symmetry representation with $\delta t\ne0$.
Momentum-dependent symmetries of this kind belong to the generalized
Noether framework~\cite{stephani,mouchet}.
Explicitly, Eq.~(\ref{rules-infini-t}) gives
\begin{align*}
\delta p_{1}&=\epsilon
\left(\onehalf p_{2}^{2}-\onehalf p_{1}^{2}+\Hv\right),&
\delta p_{2}&=-\epsilon\,p_{1}p_{2},\\
\delta q_{1}&=\epsilon\left(q_{1}p_{1}+q_{2}p_{2}\right),&
\delta q_{2}&=\epsilon\left(p_{1}q_{2}-p_{2}q_{1}\right),\\
\delta\Hv&=0,&\delta t&=\epsilon\,q_{1}.
\end{align*}
The boundary function in the action variation is
\begin{displaymath}
B=\onehalf q_{1}p_{1}^{2}+q_{2}p_{1}p_{2}-\onehalf q_{1}p_{2}^{2}.
\end{displaymath}
Thus the action is preserved up to $\epsilon[B]_{s_{0}}^{s_{1}}$.
The parametrization is consistent with the transformation because
\begin{displaymath}
\delta r=\epsilon p_{1}r,\qquad
\dfrac{T}{s}=\frac{r}{\ell}
+\epsilon\dfrac{q_{1}}{s}
=\frac{r}{\ell}(1+\epsilon p_{1})
=\frac{R}{\ell}+\mathcal O(\epsilon^{2})
\end{displaymath}
on the physical hypersurface, where $R=\sqrt{Q_{1}^{2}+Q_{2}^{2}}$.

The infinitesimal rules for the configuration variables depend
linearly on $q_{1},q_{2}$, with coefficients depending on the momenta.
In matrix form,
\begin{displaymath}
{\left.\begin{pmatrix}Q_{1}\\ Q_{2}\end{pmatrix}\right|}_{T}=
\left[\Eins+\AB_{\epsilon}\right]
{\left.\begin{pmatrix}q_{1}\\ q_{2}\end{pmatrix}\right|}_{t}
+\mathcal O(\epsilon^{2}),\qquad
\AB_{\epsilon}=\epsilon{\left.\begin{pmatrix}
\hphantom{-}p_{1}&p_{2}\\ -p_{2}&p_{1}\end{pmatrix}\right|}_{t},
\end{displaymath}
where $\Eins$ is the $2\times2$ unit matrix.
To first order, this is equivalently
\begin{displaymath}
{\left.\begin{pmatrix}Q_{1}\\ Q_{2}\end{pmatrix}\right|}_{t+\delta t}=
\e^{\delta\phi}\begin{pmatrix}\hphantom{-}\cos\delta\psi&\sin\delta\psi\\
-\sin\delta\psi&\cos\delta\psi\end{pmatrix}
{\left.\begin{pmatrix}q_{1}\\ q_{2}\end{pmatrix}\right|}_{t}
+\mathcal O(\epsilon^{2}),
\end{displaymath}
with $\delta t=q_{1}\epsilon$, $\delta\phi=p_{1}\epsilon$,
and $\delta\psi=p_{2}\epsilon$.
The configuration variables at time $T=t+\delta t$ are therefore
related to those at time $t$ by an \emph{infinitesimal scaled rotation}
whose coefficients depend on the phase-space point.
The parameter $\epsilon$ remains a single constant.
The displayed matrix exponential is not an exact finite symmetry
with the initial momenta held fixed, since the momenta themselves
change along the symmetry flow.

Finally, the two representatives describe the same physical symmetry
when compared at the same time.
Their difference $I-I_{0}=q_{1}C$ generates, on $\Sigma$, a displacement
along the physical trajectory.
Subtracting this time displacement from the configuration variations gives
\begin{displaymath}
\delta q_{1}-\dfrac{q_{1}}{t}\delta t=\epsilon q_{2}p_{2},\qquad
\delta q_{2}-\dfrac{q_{2}}{t}\delta t
=\epsilon(q_{2}p_{1}-2q_{1}p_{2}),
\end{displaymath}
which are precisely the variations generated by $I_{0}$.
The same relation holds for the momentum variations.
The time-shifting expression is therefore an alternative extended
representative of the Runge-Lenz symmetry.

\section{Conclusions}
Parameterizing the action functional in terms of a system evolution
parameter $s$ enables us to put $t=t(s)$ and $q^{i}=q^{i}(s)$
on equal footing.
The extended Hamiltonian description retains its canonical form.
For $H_{\e}=N(H-\Hv)$ with $N>0$, the physical initial
condition $H-\Hv=0$ is preserved by the extended equations.
On this hypersurface, $\dd t/\dd s=N$, and the conventional
canonical equations follow in physical time.
For $N=1$, one may choose $t=s$.
Conventional canonical transformations constitute a subgroup of
extended canonical transformations.

The extended formalism permits canonical transformations that also
map time and energy.
It thereby provides a direct Hamiltonian formulation of Noether's
theorem and its local inverse.
A smooth generator commuting with the specified extended Hamiltonian
produces a local canonical symmetry of its action up to a boundary term.
Conversely, a local variational symmetry in the stated class has a
conserved canonical generator.
Every smooth conventional first integral has such a representation
when momentum-dependent transformations and boundary terms are admitted.
These conclusions require the distinction between a canonical change
of variables and a symmetry of the same Hamiltonian.
They also require the distinction between identities on the full
extended phase space and equalities restricted to physical trajectories.

The Runge-Lenz example illustrates both points.
Its energy-dependent representative differs from the conventional
invariant by a multiple of $H-\Hv$.
With $N=r/\ell$, it generates a symmetry of the fixed extended action
whose configuration-space part is an infinitesimal scaled rotation
accompanied by a time shift.
At equal physical time, this transformation agrees with the symmetry
generated by the conventional Runge-Lenz invariant.
The result is a local Hamiltonian Noether correspondence within the
stated class, without a claim of global completeness or a restriction
to configuration-space point symmetries.

\end{document}